\documentstyle[12pt,preprint,aps,epsf,floats,amssymb]{revtex}
\def\NPB#1#2#3{Nucl. Phys. B\,{\bf #1}, #3 (19#2)}

\def\PLB#1#2#3{Phys. Lett. B\,{\bf #1}, #3 (19#2)}
\def\PLBMM#1#2#3{Phys. Lett. B\,{\bf #1}, #3 (20#2)}

\def\PRD#1#2#3{Phys. Rev. D\,{\bf #1}, #3 (19#2)}
\def\PRDMM#1#2#3{Phys. Rev. D\,{\bf #1}, #3 (20#2)}
\def\PRL#1#2#3{Phys. Rev. Lett. {\bf#1}, #3 (19#2)}
\def\PRLMM#1#2#3{Phys. Rev. Lett. {\bf#1}, #3 (20#2)}

\def\dis{\displaystyle}
\def\STU{(S,\,T\,U)}
\def\ov{\overline}

\def\to{\rightarrow}

\def\dis{\displaystyle}
\def\f{\frac}
\def\[{\left[}
\def\]{\right]}
\def\({\left(}
\def\){\right)}

\def\ov{\overline}

\def\U1em{{U(1)_{\rm em}}}
\def\to{\rightarrow}

\def\STU{{(S,T,U)}}
\def\mHsm{{m_H^{\rm sm}}}
\def\95CL{{95\%\,{\rm C.L.}}}
\def\CO{{\cal O}}

\def\End{\end{document}}

\newcommand{\mrnu}{{\nu}^{\prime}}

\tighten
\begin{document}
\preprint{
\noindent
\begin{minipage}[t]{3in}
\begin{flushleft}
\end{flushleft}
\end{minipage}
\hfill
\begin{minipage}[t]{3in}
\begin{flushright} 
CALT-68-2313\\
MIT--CTP--3080\\
UT-HEP-01-018\\
hep-ph/0102144\\
\vspace*{.7in}
\end{flushright}
\end{minipage}
}

\draft

\title{
Extra Families, Higgs Spectrum and Oblique Corrections
}

\author{{\sc Hong-Jian He}$\,^{a}$,~~
        {\sc Nir Polonsky}$\,^{b}$~~  and~~ 
        {\sc Shufang Su}$\,^{c}$  
\vspace*{.2in}
}
\address{$^{a}$ The University of Texas at Austin, Austin, Texas 78712} 
\address{
$^{b}$ Massachusetts Institute of Technology, Cambridge,  Massachusetts 02139}
\address{$^{c}$ California Institute of Technology, Pasadena, California 91125
\vspace*{.2in}}


\maketitle

\begin{abstract}
The standard model accommodates, but does not explain,  three families
of leptons and quarks, while various extensions  suggest extra matter
families.  The oblique corrections from extra chiral families  with
relatively light (weak-scale) masses, $M_{f} \sim \langle H \rangle $,
are analyzed and used to constrain the number of  extra families and
their spectrum.  The analysis is motivated, in part, by recent $N = 2$
supersymmetry constructions, but is performed in a model-independent
way. It is shown that the  correlations among the contributions to the
three oblique parameters, rather than the contribution to a particular
one, provide the most significant bound.  Nevertheless, a single extra
chiral family  with a constrained spectrum is found to be consistent
with precision data without requiring any  other new physics source.
Models with three additional families may also be accommodated but
only by invoking additional new physics, most notably,  a
two-Higgs-doublet extension.  The interplay between the spectra of the
extra fermions and the Higgs boson(s)  is analyzed in the case of
either one or two Higgs doublets, and its implications are
explored. In particular, the precision bound on the SM-like Higgs
boson mass is shown to be significantly relaxed in the presence of an
extra relatively light chiral family.
\end{abstract}

\pacs{PACS numbers:12.15.Lk, 12.60.Fr, 12.60.Jv}

\section{Introduction}
\label{sec:intro}

The number of fermion generations
is one of the unresolved puzzles within
the Standard Model (SM) of electroweak and strong interactions.
However, certain  extensions of the standard model suggest 
particular family structures.

$N=2$ supersymmetry constructions \cite{N2old,N2new},
for instance,
enforce an even number of generations,
which in practice implies three additional mirror families of 
chiral fermions (and sfermions)
with fermion masses at the weak scale,
$M_{f} \sim \langle H \rangle $,
where $\langle H \rangle \simeq 174$\,GeV is the 
Higgs vacuum expectation value (VEV) responsible for 
the electroweak symmetry breaking.
All fermion masses in $N=2$ supersymmetry
originate at low-energy from effective 
Yukawa couplings, as shown in Ref.\,\cite{N2new}, and are chiral.
(Although the matter fermions are vector-like
in the $N=2$ limit, 
gauge invariant mass terms are forbidden by a $Z_{2}$ mirror parity 
\cite{N2old,N2new}.)
The mirror fermion spectrum is bounded from
above by requiring perturbativity, and from below by 
direct collider searches. 
Hence, the natural mass range for the mirror fermions is roughly,
\begin{eqnarray}
~\dis\f{m_Z}{2} ~\lesssim~ M_f ~\lesssim~  \CO(\langle H \rangle ) ~,
\label{eq:mirrorF-MR}
\end{eqnarray}
where $m_Z \simeq 91.19$\,GeV is the mass of the weak gauge boson $Z^0$.
Here, the generic lower bound is given by the LEP $Z$-decays to heavy neutrinos
and other charged fermions. The current direct bound on
charged heavy leptons
is about $100$\,GeV, while extra SM-like quarks $(t',b')$
should be heavier than $\sim\!100-200$\,GeV, depending on detailed assumptions
regarding their mixing with $(t,\,b)$ and their decay modes of
$t'\to b+W$ and $b'\to b+Z$, etc \cite{data}. 
For simplicity, we assume hereafter
no mixing of the extra fermions among themselves and with the SM fermions
(as the latter is suppressed by the mirror parity in $N=2$), 
and in particular, that the mass range  (\ref{eq:mirrorF-MR}) 
would apply.

Eq.~(\ref{eq:mirrorF-MR}) provides a restrictive range 
which is quite different from the case of 
dynamical symmetry breaking scenarios, 
such as technicolor, where the strongly interacting
techni-fermions are generally heavy, with masses around 
or above the TeV scale \cite{TC,TCF,STU}. 
The quantum oblique corrections, parameterized in terms of the 
$S$, $T$ and $U$ parameters \cite{STU}, 
are extracted from the electroweak precision data\,\cite{data,Osaka} 
and are known to exclude such 
extra heavy chiral-fermion generations\,\cite{EP}.
For instance, one extra SM-like heavy family
would contribute to the $S$-parameter by an amount of
\begin{eqnarray}
\Delta S&=& \dis\f{1}{3\pi}\sum_j N_{cj}[I_{3L}(j)-I_{3R}(j)]^2 
  =\f{2}{3\pi}\simeq 0.21 \,,
\label{eq:DHF-S}
\end{eqnarray}
in the degenerate limit\,\cite{STU,EP}, where $I_{3L,R}(j)$ 
is the third component of weak-isospin
of the left (right) handed fermion $j$, and $N_{cj}=
3\,(1)$ denotes the color number of quarks (leptons).  
On the other hand, a nondegenerate heavy fermion doublet
$(\psi_1,\,\psi_2)$ with masses $(M_1,M_2)$ 
can yield a sizable positive $T$ which, in the limit
$|M_1-M_2|\ll M_{1,2}$, reads\,\cite{STU,veltman} 
\begin{eqnarray}
\Delta T&\simeq& \dis\f{N_{cj}}{12\pi s_W^2c_W^2}\(\f{M_1-M_2}{m_Z}\)^2 \,,
\label{eq:NDHF-T}
\end{eqnarray}
where $s_W=\sin\theta_W$ 
with $\theta_W$ being the weak angle.
Such nondecoupling effects of heavy chiral fermions
are due to the dependence
of their masses on the Yukawa couplings,
that necessarily violates the decoupling theorem\,\cite{Decouple}.
The heavy (chiral) fermion corrections (\ref{eq:DHF-S}) 
and (\ref{eq:NDHF-T}) are inconsistent with electroweak data
(when considered separately), and are often the basis for ruling out
such heavy fermion scenarios\,\cite{EP}.  
(This is contrary to the case of vector-like
fermions whose contributions to all oblique parameters decouple as $1/M^2$
and which play a crucial role, for instance, in the recent top-quark seesaw
models with either vector singlet\,\cite{DH} or doublet\,\cite{He} 
heavy fermions.)

One expects models with relatively light extra chiral fermions
to also receive non-trivial constraints from
the electroweak quantum corrections,
though the nature of the constraints may be very different.
In this work, we study the oblique corrections from the 
such relatively light new fermions [cf. eq.\,(\ref{eq:mirrorF-MR})],
as well as from the Higgs sector which generates the chiral fermion masses.
Since the extra fermions under consideration are relatively light, 
they can have a sizable mass-splitting, such as
$|M_1-M_2|\sim m_Z \not\ll M_{1,2}$, without 
causing an unacceptably large $T$. 
At the same time, the 
$S$-parameter may receive additional negative corrections.
Interestingly, a single relatively heavy SM Higgs boson leads to a sizable
negative contribution to $T$, 
and thus allows for a larger isospin breaking in the fermion
sector. For one extra fermion family with a proper spectrum, 
a SM Higgs boson as heavy as 500\,GeV is found to be 
consistent with the precision electroweak data.
Such an interplay is nontrivial, and as we will show, 
in order to accommodate up to three new families, an extended 
Higgs sector with two Higgs doublets 
(and with a highly constrained spectrum)
has to be considered.

We begin, in Sec.\,\ref{sec:oblique1},
with a summary of the definitions of the oblique parameters
$\STU$  and their current experimental bounds, 
and examine in detail the contributions in the extra lepton-quark
sector and the two-Higgs-doublet sector.
We study the interplay between the fermion and Higgs sectors in
Sec.\,\ref{sec:N2}, where $\STU$ bounds are imposed for 
deriving the allowed parameter space.
This is done first in the simplest
case with a single extra fermion family and the one Higgs doublet,
and then in the case with three extra fermion families and
the two Higgs doublets. Low energy $N=2$ supersymmetry, 
which provides an explicit theoretical 
framework in the latter case, is briefly reviewed as well.
We conclude in Sec.\,\ref{sec:sum}.
The Appendix summarizes the complete formulae for the two-Higgs-doublet 
contributions to $\STU$.

\section{New Physics Corrections to Oblique Parameters}
\label{sec:oblique1}

\subsection{The Oblique Parameters and Current Bounds}

The oblique $\STU$ parameters \cite{STU}
can be defined as  
\begin{eqnarray}
S&=&-16\pi\frac{\Pi_{3Y}(m_Z^2)-\Pi_{3Y}(0)}{m_Z^2} \,,
\label{eq:S}\\[2mm]
T&=&4\pi\frac{\Pi_{11}(0)-\Pi_{33}(0)}{s_W^2c_W^2m_Z^2} \,,
\label{eq:T}\\[3mm]
U&=&16\pi\frac{[\Pi_{11}(m_Z^2)-\Pi_{11}(0)]
              -[\Pi_{33}(m_Z^2)-\Pi_{33}(0)]}
{m_Z^2} \,,
\label{eq:U}
\end{eqnarray}
where the weak-mixing angle $\theta_W$ is defined at the scale $\mu =m_Z$.
In eqs.~(\ref{eq:S})-(\ref{eq:U}),
$\Pi_{11}$ and  $\Pi_{33}$
are the vacuum polarizations of isospin currents, and 
$\Pi_{3Y}$ the vacuum polarization of one isospin and 
one hypercharge currents. 
The above definitions\footnote{
The $\STU$ definitions used in Ref.\,\cite{EP} are equivalent to the
above eqs.\,(\ref{eq:S})-(\ref{eq:U})
though the former are defined in term of the gauge boson
mass eigenstates instead of the weak eigenstates.}
slightly differ from the original ones \cite{STU} for $(S,U)$ since we 
use the differences of $\Pi$-functions rather than their first derivatives
(with higher powers of $q^2=m_Z^2$ truncated).
Eqs.\,(\ref{eq:S})-(\ref{eq:U}) are
more appropriate for our current analysis
in which the scale of the relevant new fermions is relatively low.
The new physics corrections to $\STU$ are defined relative to their
SM reference point and are often denoted by
$(S_{\rm new},\,T_{\rm new},\,U_{\rm new})$.
To simplify the notation, we will omit these subscripts hereafter.

In certain cases, three additional oblique parameters 
$(V,\,W,\,X)$\,\cite{VWX}, 
which are generally less visible, may be further included in fitting the
data. This more elaborated procedure is beyond the scope 
of the current work and 
is not expected to affect our main conclusions.
[The contributions of the new fermions to $(V,\,W,\,X)$ drop quickly as
their masses increase beyond the $Z$-pole and become 
well below the dominant oblique corections \cite{VWX}.] 
Also, the absence of mixings between new fermions and the SM
fermions implies no extra flavor-dependent vertex corrections
to the fermionic $Z$-decay width, 
which makes the oblique corrections
sufficient for describing the new physics in our case.

\begin{figure}[t]
\begin{center}
\epsfxsize= 10 cm
\leavevmode
\epsfbox[92 157 538 579]{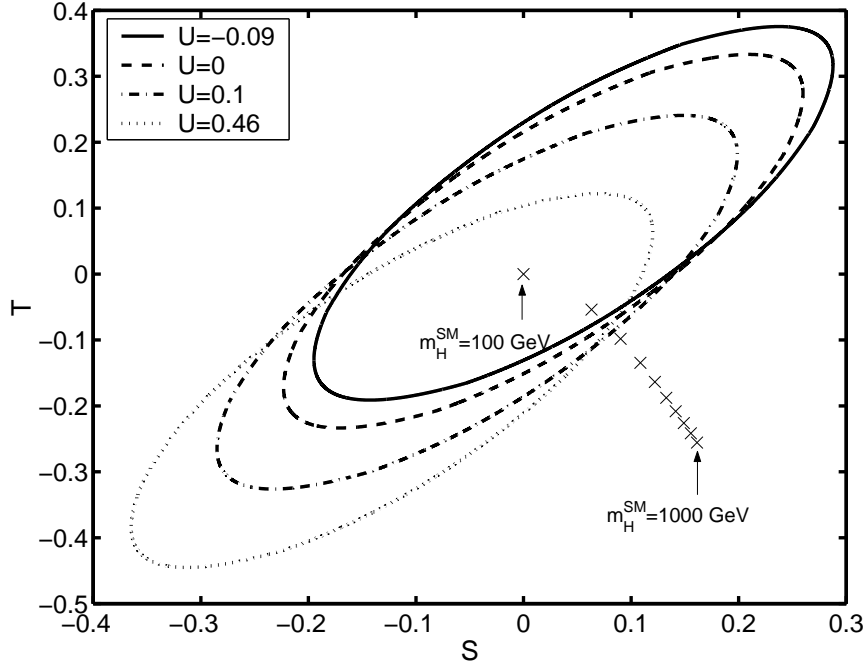}
\end{center}
\caption[f1]{The $\95CL$ contours for $S$ and $T$ 
for fixed values of $U$  (within 2$\sigma$ range) and the reference
point  $\mHsm=100$\,GeV.  The ``$\times$'' symbols denote
the SM Higgs contributions to $(S,\,T)$ 
for $\mHsm=100,\, 200,\, \dots ,\, 1000$\,GeV (from left to right)
relative to the reference point (the origin of the $S-T$ plane).
} 
\label{fig:stuexp}
\end{figure}

The updated global fit of $\STU$ to 
the various precisely measured electroweak observables
(such as the gauge boson masses $(m_Z,m_W)$, the $Z$-width $\Gamma_Z$, 
and the $Z$-pole asymmetries, etc) \cite{data,Osaka} gives\footnote{
Our global fit analysis is based on the GAPP package in Ref.\,\cite{erler1},
including the data update reported
in Ref.\,\cite{Osaka}. The newest update in Ref.\,\cite{Moriond} 
has no significant effect on our fit
and thus does not affect our conclusions.
}:
\begin{eqnarray}
S&=&   -0.04\pm{0.11}\, (-0.09)    \,,   \nonumber \\
T&=&   -0.03\pm{0.13}\, (+0.09)    \,,   \label{eq:STU-fit} \\
U&=&\ \ \, 
        0.18\pm{0.14}\, (+0.01)    \,,   \nonumber
\end{eqnarray}  
where the central values correspond to the SM Higgs mass  reference point,
$\mHsm=100$\,GeV, while the values given in the parentheses show
the changes for $\mHsm=300$ GeV.  
The uncertainties in (\ref{eq:STU-fit}) are from the inputs.
The $S$ and $T$ parameters 
are strongly correlated  as shown in the $95\%$\,C.L.
contours of Fig.\,\ref{fig:stuexp}.
Variations in $U$ mainly shift the $S-T$ contour without affecting
its shape and direction, and
a larger positive $U$ tends to diminish the allowed regions of
positive $(S,\,T)$. 

The ``$\times$'' symbols in Fig.\,\ref{fig:stuexp}
represent in SM Higgs contributions to $S$ and $T$ for 
different $\mHsm$ values relative to $\mHsm =100$ GeV.
$U$ is insensitive to $\mHsm$ for $\mHsm\gtrsim 200$\,GeV.
An important feature of the SM Higgs corrections 
is that as  $\mHsm$ increases,
$S$ becomes more positive while $T$ is driven to more negative values.
As such, a SM Higgs with a mass $\mHsm \gtrsim 300$\,GeV is clearly
outside the $95\%$\,C.L. $S-T$ contours for wide range of $U$ values.\footnote{
The best fit for a pure  SM Higgs boson  with $\STU=0$ 
gives a similar but somewhat stronger
bound,  $34\,{\rm GeV}\,\leq m_{H}^{\rm sm} \,\leq 202$\,GeV,
at 95$\%$\,C.L.} 
However,  including certain types of new physics contributions to $\STU$ may 
drastically relax the upper bound on the Higgs mass, as long as the new
corrections either 
             { \it(i)  decrease $S$}, or
            {\it (ii)  lift up $T$,}  or  
            {\it (iii) achieve both}.
As we will show in the following sections, the extra fermions under 
consideration generally lead to a large positive $T$,
and in many cases also to a sizable $S>0$. Hence, our analysis will  
mainly fall under Case {\it (ii)}.

\subsection{Lepton and Quark Sector}

For generality, we consider two fermions  $(\psi_1,\psi_2)$, 
with masses $(M_1, M_2)$  and the following SM charges, 
\begin{equation}
\begin{array}{lccc}
{\rm Fermions:} & 
\displaystyle 
\psi_L = \left( 
\begin{array}{c} 
\psi_{1L}\\ \psi_{2L} 
\end{array}             \right)\,,~~~~ 
& \psi_{1R}\,,~~~~ & \psi_{2R}\,;\\[3mm]
{\rm Hypercharge:} & Y\,,~~~~ 
& 
\displaystyle 
Y+\frac{1}{2},~~~~ & \displaystyle Y-\frac{1}{2};
\end{array}
\label{eq:charge}
\end{equation}
where the electric charge is given by
$Q_j = I_{3j}+Y_j$ with $I_{3j}$ and $Y_j$ 
being the third component of weak-isospin and the hypercharge of the
fermion $j$, respectively.  
For SM fermions, one has
$Y=\f{1}{6}\,\(-\f{1}{2}\)$ 
in eq.~(\ref{eq:charge}) for quarks (leptons).
For mirror fermions in the Minimal 
$N=2$ Supersymmetric SM (MN2SSM) \cite{N2new}, 
one has $Y=-\f{1}{6}\,\(\f{1}{2}\)$ in eq.~(\ref{eq:charge})
for mirror quarks (mirror leptons).  
(For a review on the MN2SSM, see Sec.~\ref{subsec:MN2SSM}.)
Hence, the correspondence with eq.\,(\ref{eq:charge}) is,
    $(M_1, M_2)\leftrightarrow{(M_{\nu}, M_\ell )}$ for leptons
and $(M_1, M_2)\leftrightarrow{(M_{\ell'}, M_{\mrnu})}$ for mirror leptons, 
and similarly for the quarks and mirror quarks.

Using eqs.\,(\ref{eq:S})-(\ref{eq:U}), we can compute 
the one-loop fermionic contributions 
to the oblique $\STU$ parameters as below,
\begin{eqnarray}
S_f&= & \frac{N_c}{6\pi} 
\left\{
2(4Y+3)x_1+2(-4Y+3)x_2-2Y\ln\frac{x_1}{x_2} 
\right.\nonumber \\[1mm]
&&\left. +\left[\left(\frac{3}{2}+2Y\right)x_1+Y\right] G(x_1)
 +\left[\left(\frac{3}{2}-2Y\right)x_2-Y\right] G(x_2)
\right\}\,,
\label{eq:Sfermion}\\[3mm]
T_f&=&\frac{N_c}{8\pi{s}_W^2c_W^2}F(x_1,x_2) \,,
\label{eq:Tfermion}\\[3mm]
U_f&=&-\frac{N_c}{2\pi}\left\{\frac{x_1+x_2}{2}-\frac{(x_1-x_2)^2}{3}
+\left[\frac{(x_1-x_2)^3}{6}-\frac{1}{2}\,\frac{x_1^2+x_2^2}{x_1-x_2}\right]
\ln\frac{x_1}{x_2}\right. \nonumber\\[1mm]
&&+\left.\frac{x_1-1}{6}f(x_1,x_1)+\frac{x_2-1}{6}f(x_2,x_2)+
\left[\frac{1}{3}-\frac{x_1+x_2}{6}-
\frac{(x_1-x_2)^2}{6}\right]f(x_1,x_2)\right\} ,
\label{eq:Ufermion}
\end{eqnarray}
where  $x_i=(M_i/m_Z)^2$ with $i=1,2$ and
the color factor $N_{c} = 3\,(1)$  for quarks (leptons).
The functions $G(x)$, $F(x_1,x_2)$, and $f(x_1,x_2)$ are defined 
by eqs.\,(\ref{eq:Gfun}), (\ref{eq:Ffun}) and  (\ref{eq:ffun}),
in the Appendix.
We observe that for a given $(M_1,\,M_2)$,
eq.\,(\ref{eq:Sfermion}) is invariant under the exchanges of
$Y\leftrightarrow{-}Y$ and $M_1\leftrightarrow{M}_2$, so that
the fermions $(\psi_1,\,\psi_2)$ and their mirrors $(\psi_2,\,\psi_1)$
have the same expression for $S$.
Therefore, we will not distinguish hereafter between a fermion and its mirror, but
simply use $(M_1,\,M_2)$ to denote $(M_N,\, M_E)$ in the (mirror) lepton sector and 
$(M_U, \,M_D)$ in the (mirror) quark sector.

It is instructive to consider the 
limit  $M_{1,2}^2\gg{m_Z^2}$, 
under which the $S$ parameter approximately reads,
\begin{equation}
S_f=\frac{N_c}{6\pi}\left[1-2Y\ln\left(\frac{M_1}{M_2}\right)^2+
\frac{1+8Y}{20}\left(\frac{m_Z}{M_1}\right)^2+
\frac{1-8Y}{20}\left(\frac{m_Z}{M_2}\right)^2
+O\left(\frac{m_Z^4}{M_i^4}\right)\right]   .
\label{eq:Sapp}
\end{equation}
If the mass splitting ~$|M_1 - M_2|/M_{1,2}$~ is small, 
then all mass-dependent terms decouple and eq.~(\ref{eq:Sapp}) reduces to
the positive constant term $N_c/6\pi$, which leads to the well-known 
result in eq.\,(\ref{eq:DHF-S}).
However, as long as $(M_1,\,M_2)$ are non-degenerate and not too large, 
additional negative corrections to the constant term  $N_c/6\pi$ may
arise, depending on the sign of hypercharge $Y$.

\begin{figure}[t]
\begin{center}
\epsfxsize= 12 cm
\leavevmode
\epsfbox[56 44 452 511]{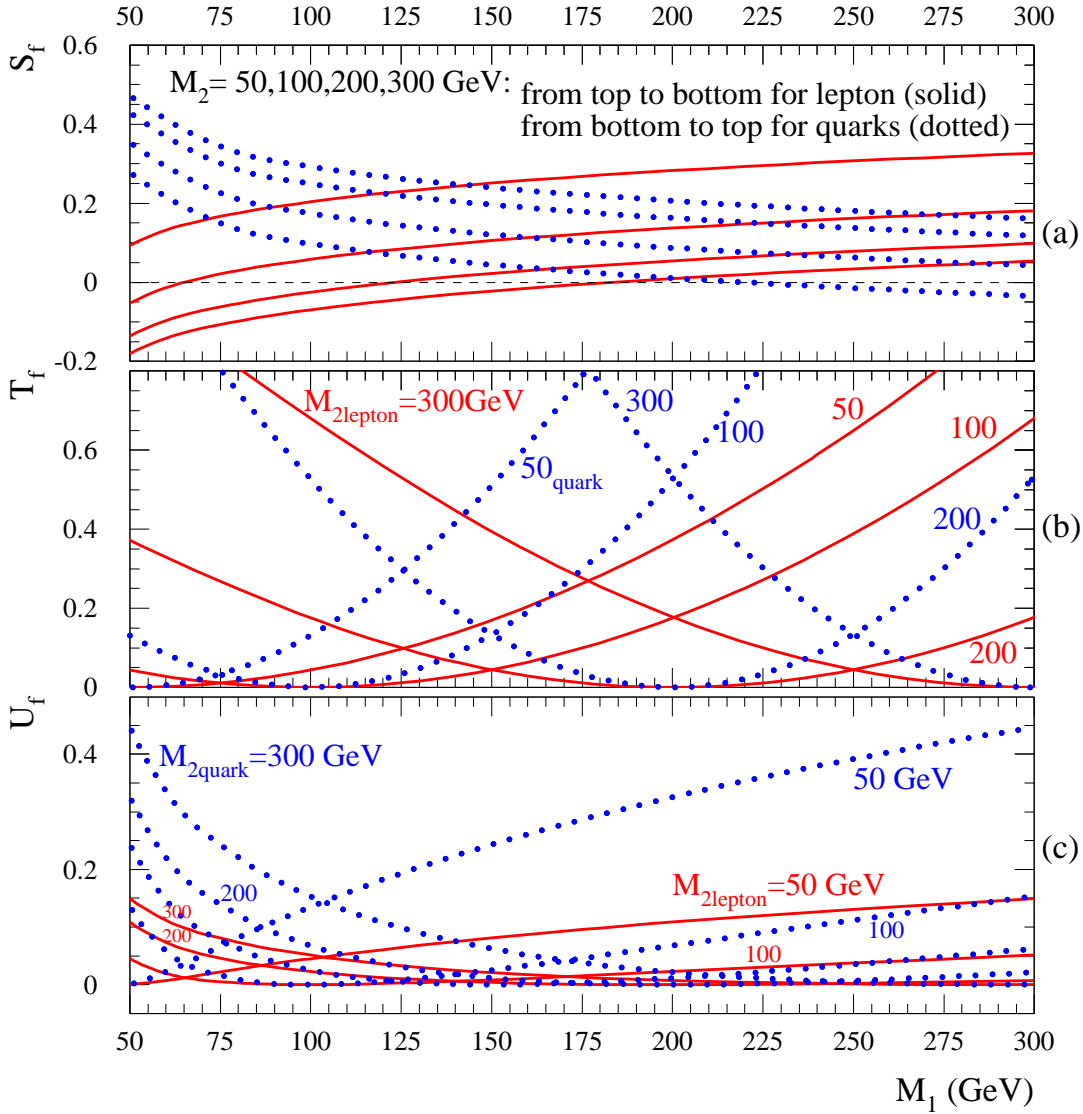}
\end{center}
\caption[f1]{The contributions to $S$, $T$ and $U$ 
from one extra family of leptons
(solid curves) and quarks (dotted curves).}
\label{fig:stufermion}
\end{figure}

The contributions to $S$ from one generation
of either ordinary or mirror leptons and quarks are shown in 
Fig.\,\ref{fig:stufermion}(a),
where the solid curves are for leptons and 
dotted curves for quarks.  
The mass range of the chiral fermions are chosen to be 
between 50 GeV and 300 GeV.  
(We note that after adding experimental bounds
on the charged extra fermions, the lower end of their mass range
would be shifted somewhat above $50$\,GeV, depending on the details
of each particular model.)
The lepton contribution to $S$ grows with 
an increasing $M_1$ ($M_N$)
and with a decreasing $M_2$ ($M_E$), while the quark contribution
behaves in the opposite way.  This is 
due to their different signs of $Y$.  The quark contribution is 
enhanced by the color factor,
but is suppressed by the smaller Y.  
For  $M_{1,2}^2 \gg{m}_Z^2, (M_1-M_2)^2$, 
$S$ should approach its asymptotic value $1/6\pi$ for leptons
and $1/2\pi$ for quarks. This may be understood from 
Fig.\,\ref{fig:stufermion}(a) by  examining 
the solid (dotted) curve with $M_2=300$\,GeV 
which already well approaches $\sim\!0.05~(0.16)$
for leptons (quarks) as $M_1$ increases to about $300$\,GeV. 
However, 
for quarks and leptons with masses $\sim\CO(m_Z)$,
smaller and  even negative values of $S$ can be obtained. Negative
values of $S$ occur in the non-degenerate region
of $M_E>M_N$ and $M_U>M_D$.  For instance, 
$(M_N,\,M_E)=(50,\,300)$\,GeV gives $S_\ell =-0.18$.

The contributions to $T$ and $U$ from chiral fermions are depicted in 
Fig.\,\ref{fig:stufermion}(b) and (c).
The parameters
$T$ and $U$ measure the weak-isospin violation in the 
${\rm SU}(2)_L$ doublet and thus are nonvanishing only for 
$M_1\neq M_2$. The more $M_1$ and $M_2$ split, the larger
their contributions to  $(T_f,\,U_f)$  become.
Furthermore, the $(T_f,\,U_f)$ formulae 
eqs.\,(\ref{eq:Tfermion}) and (\ref{eq:Ufermion}) 
are invariant under the exchange $M_1\leftrightarrow{M}_2$ and are
always positive, unlike the contributions of the  Higgs boson
(cf., Fig.\,1). 
While ${U}_f$ is relatively small, ${T}_\ell$, for example, 
could be as large as 0.68 for  
$(M_N,\, M_E)=(100,300)$ GeV.  
Since $(T_f,\,U_f)$ depend only on isospin-breaking
and are symmetric under $M_1\leftrightarrow M_2$,
their $M_{1,2}$-dependence  is the same 
for fermions and mirror fermions.
The quark contributions to $(T_f,\,U_f)$
are again enhanced by their color factor. 

In order to accommodate new fermion families,
the up- and down-type (mirror) quarks have to be sufficiently degenerate 
to avoid a too large positive $T$. 
Unfortunately, this renders $S$ positive in most of 
the parameter space.  A non-degenerate pair of (mirror) leptons 
could help to satisfy the $S$ constraint, but it also 
contributes positively to $T$ (though more moderately comparing to quark).
A positive contribution to $U$ can better fit  
the data, but it is numerically less significant, as shown in
Fig.\,\ref{fig:stufermion}(c).
Clearly, the nontrivial correlations among lepton and quark contributions
to all three oblique parameters (rather than to any particular one)
provide the most significant constraints.

\begin{figure}[t]
\begin{center}
\epsfxsize= 10 cm
\leavevmode
\epsfbox[93 187 539 579]{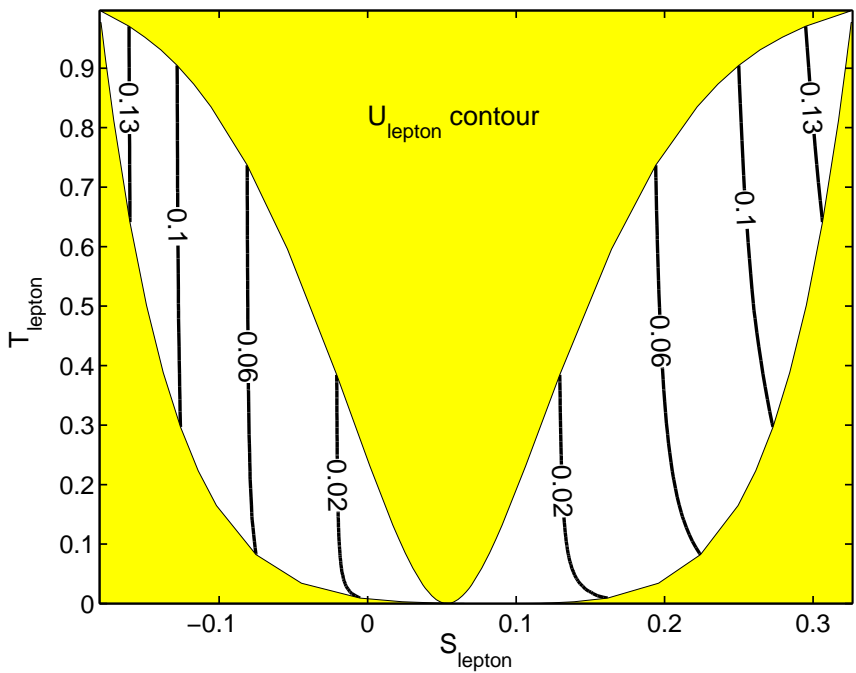}
\end{center}
\caption[f1]{$U$ contours in the
$S-T$ plane for one generation of (mirror) leptons, 
which are derived from 
eqs.\,(\ref{eq:Sfermion})-(\ref{eq:Ufermion})
for the mass range $50 \leq m_{\ell} \leq 300$\,GeV
and with no experimental bounds imposed.
The shaded areas cannot be theoretically reached.}
\label{fig:stulepton}
\end{figure}

\begin{figure}[t]
\begin{center}
\epsfxsize= 10 cm
\leavevmode
\epsfbox[93 187 539 579]{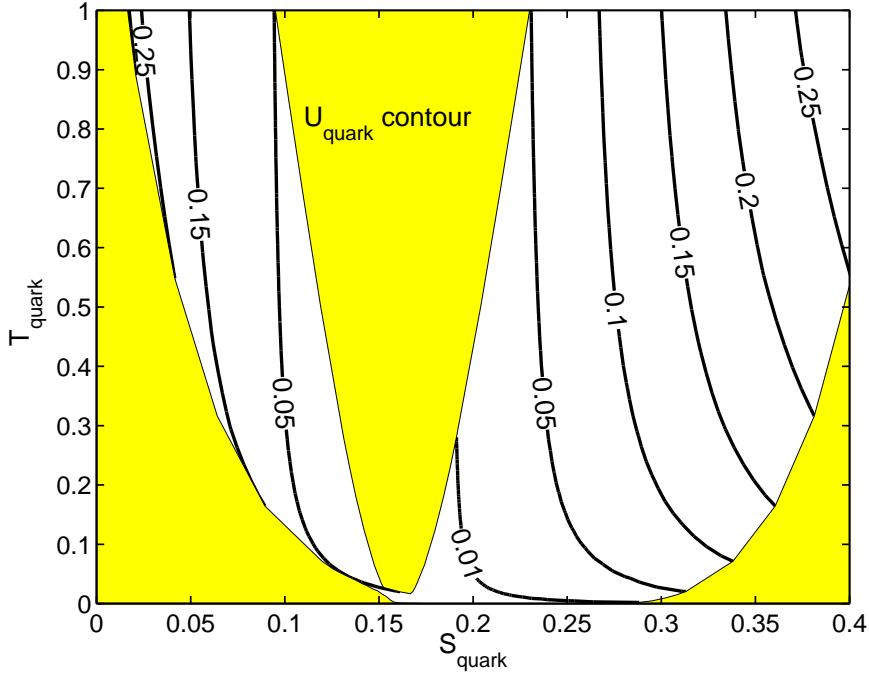}
\end{center}
\caption[f1]{$U$ contours in the
$S-T$ plane for one generation of (mirror) quarks, 
which are derived from 
eqs.\,(\ref{eq:Sfermion})-(\ref{eq:Ufermion})
for the mass range $50 \leq m_{q} \leq 300$\,GeV
and with no experimental bounds imposed.
The shaded areas cannot be theoretically reached.}
\label{fig:stuquark}
\end{figure}

In order to compare the theoretical predictions with the current 
experimental constraints 
shown in Fig.\,\ref{fig:stuexp},
it is very instructive to depict the above fermionic oblique corrections 
eqs.~(\ref{eq:Sfermion})-(\ref{eq:Ufermion}) in the
$S-T$ plane for given values of $U$.  This corresponds to a set of 
``$U$-contours''  in the theoretically allowed regions of the $S-T$ plane,
which should be directly compared to the experimental bounds 
of Fig.\,\ref{fig:stuexp}.
In Figs.\,\ref{fig:stulepton} and \ref{fig:stuquark}, we plot various
$U$-contours in the $S-T$ plane for one family of leptons and of quarks, 
respectively.   
For leptons, $S_\ell$ can be negative in large regions of the parameter space.
For quarks,  $S_q>0$ in most of the parameter space as
to avoid a too large contribution to $T_q$.
Although a positive $U_{f}$ is consistent with the data, 
$T_{f}$ provides a very strong constraint when combined with  $S_{f}$.  
Nevertheless, comparing with the $S-T$ fits in Fig.\,\ref{fig:stuexp},
one finds that one extra chiral family is viable, even without additional 
new physics contributions.
This is consistent with the recent study in 
Ref.\,\cite{previous}, where a similar conclusion was reached.
Ref.\,\cite{previous} used an unconventional formalism for 
analyzing the oblique corrections and
a detailed comparison is difficult.
Our analysis, based on the standard $\STU$ formalism\,\cite{STU},
is transparent and can be readily applied to a given model.
In what follows, we focus on the interplay
between extra families and the Higgs sector. We aim at accommodating
up to three chiral families (as theoretically motivated by our recent
$N=2$ constructions \cite{N2new}), which requires to extend the Higgs
sector with two-doublets. 
Henceforth, our study substantially differs from Ref.\,\cite{previous}.

Finally, we note that it should be straightforward to translate above
Figs.\,\ref{fig:stulepton} and \ref{fig:stuquark} to any 
number $N_{g}$ of extra generations, 
i.e., for $N_g>1$, the same curves represent 
the oblique parameters with the values $\STU/N_g$,
if one assumes that these new generations are degenerate  
in mass with each other.
However, it is extremely difficult
to accommodate more than one extra generation 
with the data.
We will return to this issue in Sec.~\ref{sec:N2}.

\subsection{Two Higgs Doublet sector}
\label{sec:2hdm}

The exact corrections to $\STU$ in a general 
two-Higgs-doublet model (2HDM) 
have been computed in Ref.~\cite{haber}. 
We will denote these contributions by $S_H$, $T_H$, and $U_H$, respectively.
Their explicit formulae are lengthy
and are summarized in the Appendix for completeness.
For $N=1$ supersymmetry, and in particular the
$N=1$ minimal supersymmetric extension of the SM (MSSM)
(with high-scale supersymmetry breaking), 
the Higgs contributions are generally small
due to the tree-level constraints among the masses 
of the light and heavy CP-even, the CP-odd, 
and the charged Higgs bosons, 
($m_h,\, m_H,\, m_A,\, m_{H^{\pm}}$, respectively).
However, for a two-Higgs-doublet sector with a general 
Higgs mass spectrum, significant contributions can arise in
large regions of the parameter space.
Such non-MSSM-like Higgs spectrum may be realized for a
$N=1$ or $N=2$ supersymmetry scenario
with a sufficiently low scale of supersymmetry breaking \cite{hard}.

\begin{figure}[t]
\begin{center}
\epsfxsize= 10 cm
\leavevmode
\epsfbox[93 187 539 579]{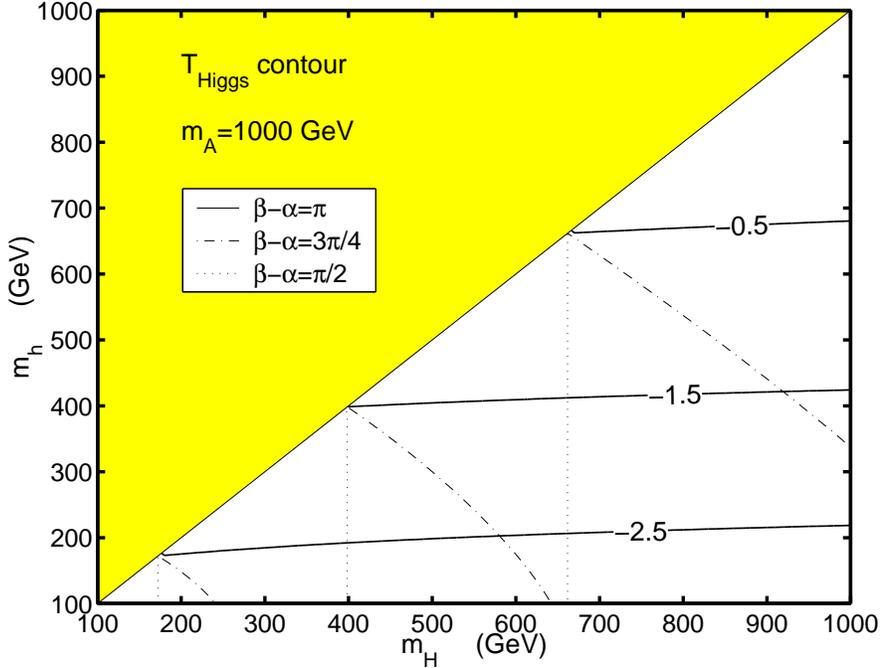}
\end{center}
\caption{Contours for $T_H$ in the 2HDM for
$m_A=1000$\,GeV and $\beta - \alpha=\pi$ (solid line), 
$\f{3\pi}{4}$ (dash-dotted line) and 
$\f{\pi}{2}$ (dotted line). 
Here we consider $\tan\beta>1$ 
($\f{\pi}{4}<\beta<\f{\pi}{2}$) and $-\f{\pi}{2}<\alpha<0$,
so that $\f{\pi}{4}<\beta-\alpha<\pi$.
The $m_{H^{\pm}}$ value is chosen to minimize $T_H$.     
These contours are derived from eq.~(\ref{eq:Thiggs}).}
\label{fig:Thiggs1000}
\end{figure}

\begin{figure}
\begin{center}
\epsfxsize= 10 cm
\leavevmode
\epsfbox[93 187 539 579]{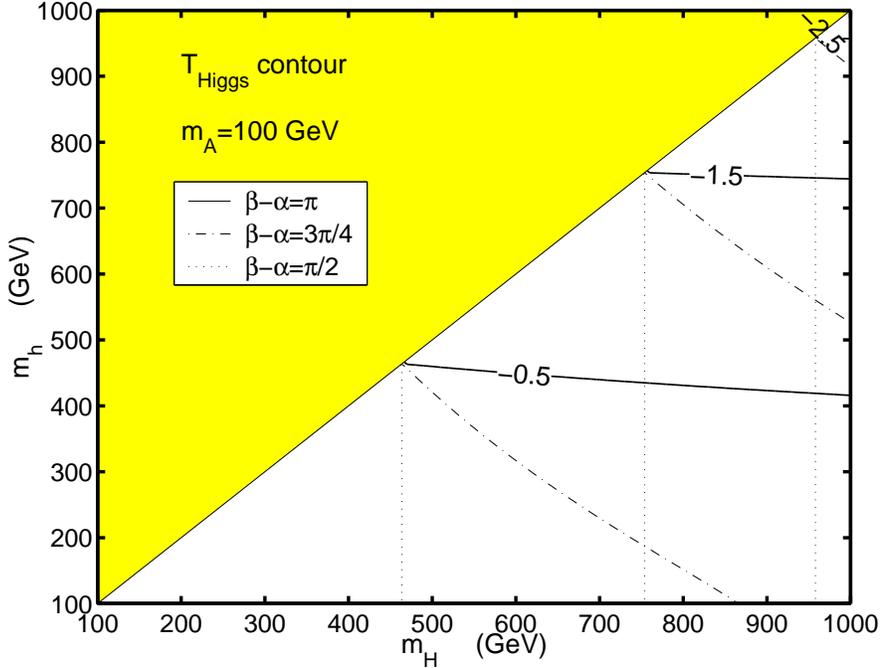}
\end{center}
\caption{Contours for $T_H$ in the 2HDM for
$m_A=100$\,GeV.  Notation is the same as in Fig.\,\ref{fig:Thiggs1000}.
}
\label{fig:Thiggs100}
\end{figure}

The contribution
$T_H$ could be either positive or negative, depending on the spectrum 
of the Higgs masses and on the difference between the two rotation angles
($\beta-\alpha$), where
$\tan\beta = \langle H_2\rangle/ \langle H_1\rangle$ 
[with $H_{1}$ ($H_{2}$) being the Higgs doublet of
 negative (positive) hypercharge] and
$\alpha$ is the rotation angle for obtaining the 
CP-even mass-eigenstates $(h^0,\,H^0)$.
The $T$-contours in the $(m_h,m_H)$ plane for 
$m_A=1000$\,GeV and $m_A=100$\,GeV are shown
in Figs.\,\ref{fig:Thiggs1000} and \ref{fig:Thiggs100} for 
$\beta-\alpha=\pi$ (solid line), $\f{3\pi}{4}$ (dash-dotted line) 
and $\f{\pi}{2}$ (dotted line), 
where $m_{H^{\pm}}$ is chosen as to minimize $T_H$.
A negative contribution to $T_H$  can always be achieved with 
an appropriately chosen $m_{H^{\pm}}$. (This was also
noted in Ref.\,\cite{grant}.) 
For some values of $m_{H^{\pm}}$,
$T_H$ could be positive and large, 
however, we will concentrate hereafter only on the 
more interesting regions with negative $T_H$.

The regions which
correspond to 
a sizable negative $T_H$ can be classified as follows: 
\begin{itemize}
\item{Large $m_A$:
(Ia) $m_h\ll m_{H^{\pm}}\ll{m}_A$,~ $\beta-\alpha\sim\pi$\,;\\[1.5mm]
\hspace*{1.89cm}
(Ib) $m_h\sim{m}_H \ll m_{H^{\pm}}\ll{m}_A$\,;}
\item{Small $m_A$:
(IIa) $m_A\ll{m}_{H^{\pm}}\ll m_H$,~ $\dis\beta-\alpha\sim\f{\pi}{2}$\,;\\[1.5mm]
\hspace*{1.89cm}
(IIb) $m_A\ll{m}_{H^{\pm}}\ll m_h\sim{m}_H$\,;}
\end{itemize}
where the minimum value for $T_H$ is achieved  
for $m_{H^{\pm}}\simeq{0.6}\ m_{\rm heavy}$ and 
$m_{\rm heavy}=\max(m_H,\,m_A)$.
This can be understood by examining the approximate 
formula for $T_H$ in the limit $m_{\rm Higgs}^2\gg{m_Z}^2$ \cite{grant}:
\begin{eqnarray}
T_H&=&\frac{1}{16\pi{s}^2_Wm^2_W}\left\{\cos^2({\beta-\alpha})
\[F(m_{H^{\pm}}^2, m_h^2)+F(m_{H^{\pm}}^2, m_A^2,)-F(m_{A}^2, m_h^2,)\]\right.
\nonumber \\
&&\hspace*{19.3mm}+\left.\sin^2({\beta-\alpha})
\[F(m_{H^{\pm}}^2, m_H^2)+F(m_{H^{\pm}}^2, m_A^2)-F(m_{A}^2, m_H^2)\]\right\},
\label{eq:thiggsapp}
\end{eqnarray}
where 
$F(x_1,x_2)$ is defined in eq.\,(\ref{eq:Ffun}).
[The approximate formulae for ($S_H,\,U_H$) are given in the Appendix for 
completeness.]   
Terms inside the first (second) bracket are symmetric in 
$m_{h(H)}$ and $m_A$, and could obtain large negative values if 
there is a large split between $m_{h(H)}$ and $m_A$ and
$m_{\rm light}\ll{m}_{H^{\pm}}\ll{m}_{\rm heavy}$.  
For $\beta-\alpha=\pi \,[{\pi\over 2}]$, we have
$\sin^2({\alpha-\beta})=0 \,[\cos^2({\alpha-\beta})=0]$,
so that only the first (second) bracket contributes, which is independent of 
$m_H$ ($m_h$).  This is the case in region (Ia) and (IIa).
For general values of $\beta-\alpha$, 
$m_h$ and $m_H$ have to be 
sufficiently close in order for $T_{H}$ to be large and negative. 
This is the case in regions (Ib) and (IIb).  We also notice that in 
Figs.~\ref{fig:Thiggs1000} and \ref{fig:Thiggs100}, each set of 
$T_H$-contours approach the same point at the boundary
of $m_h = m_H$.  This is because the dependence on $\beta - \alpha$
disappears under this limit 
[see eqs.~(\ref{eq:Thiggs}) and (\ref{eq:thiggsapp})].

We note that the parameter $T_H$ can be as negative as
$-2.5$, and could cancel
large positive contributions from the quark and lepton sector
when more than one extra family is included.
$S_H$ and $U_H$ are relatively small in these two regions, where one has
an almost positive $S_H<0.1$
and a negative $U_H$ with $|U_H|<0.02$.
In Case-(Ia), a sizable positive
$S_H\sim{0.16}$ and
a slightly negative $U_H\sim{-0.05}$ are also possible.

Clearly, the Higgs spectrum in these two regions is very different from that of
the conventional $N = 1$ MSSM. Even in the case
of a more general supersymmetry breaking 
scenario\,\cite{hard},
it requires some fine-tuning of the mass parameters and
the quartic couplings.  In principle,  such relations are easier 
to realize in models with more than two Higgs doublets (such as
$N=2$ supersymmetry), where more Higgs states can exist 
at the scale $m_{\rm heavy}$ or above
and thus  considerably expand the parameter space.

The correlations between the spectra of the minimal
one- or two-Higgs-doublet sector and the
additional chiral families via the precision $\STU$ constraints  
will be systematically analyzed in the next section.

\subsection{Other Super and Mirror Particles}

The contributions of the $N=1$ sparticles, 
with a typical mass scale $M_{\rm SUSY}$, 
to the oblique parameters are generally
small in the decoupling region $M_{\rm SUSY}\gg{m}_Z, m_t$,
which we will assume in our analysis for simplicity. 
In practice, this only requires $M_{\rm SUSY} \gtrsim 300$ GeV, as shown
in Refs.~\cite{haber,stop,sparticle}. 
Aside from sfermions and mirror sfermions,
there could also be visible contributions from Majorana fermions, 
such as gauginos, Higgsinos, and, in $N=2$, mirror gauginos
and Higgsinos. In general, contributions from
Majorana fermions to $S$ could have either sign \cite{previous,majorana}.

In our current study we  concentrate on the
contributions of the Higgs bosons and of (mirror) quarks and leptons.
For simplicity, the effects from  sfermions and Majorana fermions 
are assumed to be negligible. This is indeed the case in the decoupling regime
  $M_{\rm SUSY}\gg{m}_{Z},m_t$ under consideration.
Clearly, an arbitrary spectrum of sparticles and/or mirror
gauginos will add  more degrees of freedom to fit the data and thus
further relax the correlations derived in the next section.
A more elaborate analysis including these complications is left for 
future work.

\section{Spectra of Extra Fermions and Higgs Bosons:\\
The Interplay
}
\label{sec:N2}

\subsection{Interplay of Extra Fermions and One-Higgs-Doublet Sector}

\begin{figure}[t]
\begin{center}
\epsfxsize= 12 cm
\leavevmode
\epsfbox[93 180 521 610]{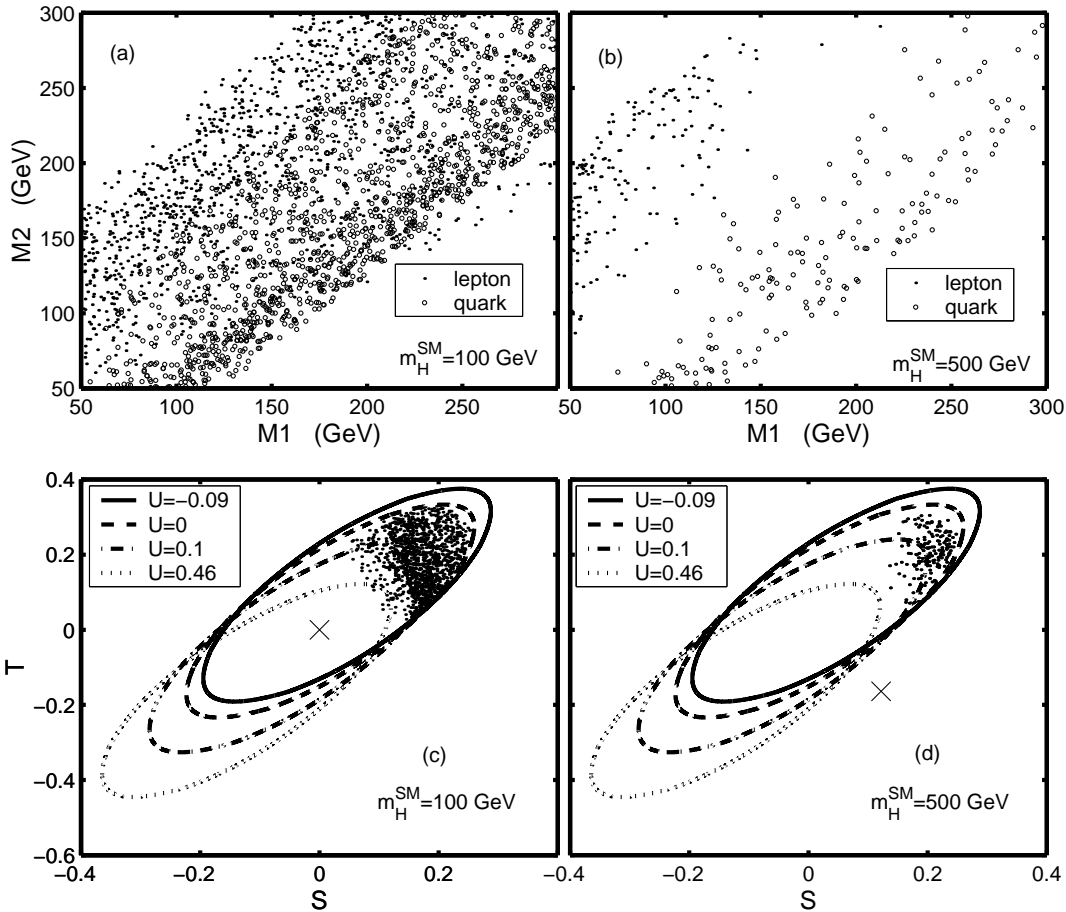}
\end{center}
\caption{
The $M_1-M_2$ plane is scanned assuming 
one extra family of leptons (dots) and 
of quarks (circles) and one Higgs doublet.
The allowed region is defined by the 
models which are admitted by the data.
Figure (a) is shown for $\mHsm =100$\,GeV and Figure (b) for $\mHsm =500$\,GeV.
In Figures (c) and (d), the allowed points are shown in
the $S-T$ plane (with the $\95CL$ experimental bounds displayed
for comparison) for $\mHsm=100$\,GeV and 500\,GeV, respectively, and
the ``$\times$'' symbols denote the corresponding 
SM Higgs contributions. 
}
\label{fig:oneg}
\end{figure}

We begin by considering
the simplest case with one extra (mirror) family 
and one (SM) Higgs doublet.
We display in Fig.\,\ref{fig:oneg} (a) and (b), 
the $M_1-M_2$ plane, 
where each point represents an 
experimentally viable four-generation model, and  
dots and circles represent leptons and quarks,
respectively.  The initial sample consists of 10000 models.
We choose, for illustration, a light SM Higgs with mass 
$\mHsm=100$\,GeV [cf., Fig.\,\ref{fig:oneg}(a)]
and a heavy SM Higgs with mass
$\mHsm=500$\,GeV [cf., Fig.\,\ref{fig:oneg}(b)].
Large regions of the parameter space are allowed, 
where the preferred regions are given by 
$M_2>M_1$ for leptons and $M_1>M_2$ for quarks.
For a heavy Higgs boson $\mHsm=500$\,GeV, the 
leptons and quarks occupy different mass regions, while
in the case of a very light Higgs boson
they largely overlap.
Future discoveries of light extra
lepton/quark spectra can provide important information about the
Higgs boson mass range, and vice versa.
Figs.\,\ref{fig:oneg}(c) and (d) display
the corresponding points in
the $S-T$ plane with the $\95CL$ experimental bounds superimposed for
$\mHsm=100$\,GeV and 500\,GeV, respectively.
From Fig.\,\ref{fig:oneg}(d), we see that for one extra chiral family, a
heavy SM Higgs with  $\mHsm=500$\,GeV can be accommodated via the
scenario of large fermionic $T_f>0$.

We note in passing that,
after the completion of this work,
Ref.~\cite{PeskinNew} analyzed the limits on a heavy SM Higgs boson
in the case of TeV-scale heavy technifermions
which generate a large positive contribution to $T$.
Our study has solely focused on relatively light extra chiral families with
masses significantly below $\CO ({\rm TeV})$, as motivated 
by $N=2$ supersymmetry constructions\,\cite{N2old,N2new}. The relaxation
of the Higgs mass limits derived
from precision electroweak data could be significant
in either case.

\subsection{Minimal $N=2$ Supersymmetric SM and Mirror Families}
\label{subsec:MN2SSM}

Before proceeding to discuss the case with three extra chiral families
and the two-Higgs-doublets,
a review of the theoretical framework which motivates this scenario
is in place.
As mentioned earlier, this spectrum arises in constructions
of low-energy $N=2$ supersymmetry.
Low-energy realizations of $N=2$ supersymmetry and its related 
phenomenology were recently investigated in Ref.\,\cite{N2new}. 
In the minimal $N=2$ supersymmetric SM, 
for each of the ordinary quark (lepton) and its
squark (slepton) superpartner of the $N=1$ extension, there is also a conjugate
{\it mirror} quark ({\it mirror} lepton)  and its 
mirror squark (mirror slepton) superpartner.
For each gauge boson and gaugino, 
there is also a {\it mirror} gauge boson and a {\it mirror} gaugino.
The Higgs and Higgsino are also accompanied by their mirrors.   
In particular, three additional mirror generations of
chiral fermions are predicted in the MN2SSM.

The mirror quarks and leptons do not obtain gauge-invariant 
vectorial mass terms (which would mix the mirror and ordinary sectors) 
due to a $Z_{2}$  mirror parity \cite{N2new}. Instead, their
masses arise from effective Yukawa interactions
and are thus proportional
to the relevant Higgs VEVs of electroweak symmetry breaking (EWSB). 
As such, their mass range is constrained to be at
the weak scale [cf., eq.\,(\ref{eq:mirrorF-MR})].
In order to realize $\CO(1)$ effective Yukawa couplings 
at low energies, supersymmetry itself is broken at
a low scale. The large Yukawa couplings also imply that
mirror fermion/sfermion loops can significantly modify
the CP-even Higgs spectrum at one loop.
(This is similar to the usual top/stop sector, but now all
three mirror families may contribute).

The MN2SSM Higgs sector 
is less constrained than that of the MSSM or other $N=1$ frameworks. 
In particular, any one of the four Higgs doublets 
which appear in MN2SSM \cite{N2new} could
participate in EWSB.
Even when assuming for simplicity a MSSM-like Higgs structure with two-doublets
participating in the EWSB, 
the $N=2$ two-Higgs-doublet spectrum could be quite different from that
of the MSSM.
This is because the
tree-level Higgs quartic couplings $\lambda$ arise not only from supersymmetric
terms  $\lambda \sim g^{2}$, for $g$ being the gauge coupling, 
as in the MSSM, but also 
from hard supersymmetry breaking operators (whose generation
goes hand in hand with that of the effective Yukawa couplings)
$\lambda_i \sim (g^{2}) + \kappa_i$ \cite{hard}, where 
$\kappa_i$ is the contribution from higher order operators in 
the K\"{a}hler potential.
Therefore, the usual MSSM relations among the Higgs mass eigenvalues 
 $m_h\ll{m}_H\sim{m}_A\sim{m}_{H^\pm}$ (assuming $m_A\gg{m}_Z$) no 
longer hold, and the physical Higgs mass spectrum is somewhat
arbitrary. 
This observation is generic to
any theory with low-energy supersymmetry breaking where 
$\kappa \sim {\cal{O}}(1)$ is realized \cite{hard}. 

We note in passing that  models with higher dimensions
often lead after compactification
to an effective $N=2$ structure in four-dimensions.
Therefore, our analysis of $N=2$ models and of the  associated mirror families 
may be applied in certain cases to theories with large extra dimensions.

\subsection{Interplay of Extra Fermions and Two-Higgs-Doublet Sector}

It was shown above
that one extra chiral generation 
($N_g = 1$) can be 
accommodated by the precision data with 
the SM Higgs mass up to about   $500$\,GeV. 
This is not the case for $N_{g} = 2$ and
$N_{g} = 3$. In fact, the $N_g =3$ case, 
as predicted in the MN2SSM, requires 
additional new physics contributions 
(beyond that of a single Higgs doublet) to the
oblique parameters. The minimal version of such an extension 
is to invoke the two-Higgs-doublet sector. 
For generality (and being consistent with
the $N=2$ framework described above),
we will consider a general 2HDM.
Thus, our analysis is valid for any given model which
contains two Higgs doublets together with extra families, and our
constraints on the parameter space can be readily applied to any such model.

\begin{figure}[h]
\begin{center}
\epsfxsize= 11 cm
\leavevmode
\epsfbox[113 180 521 610]{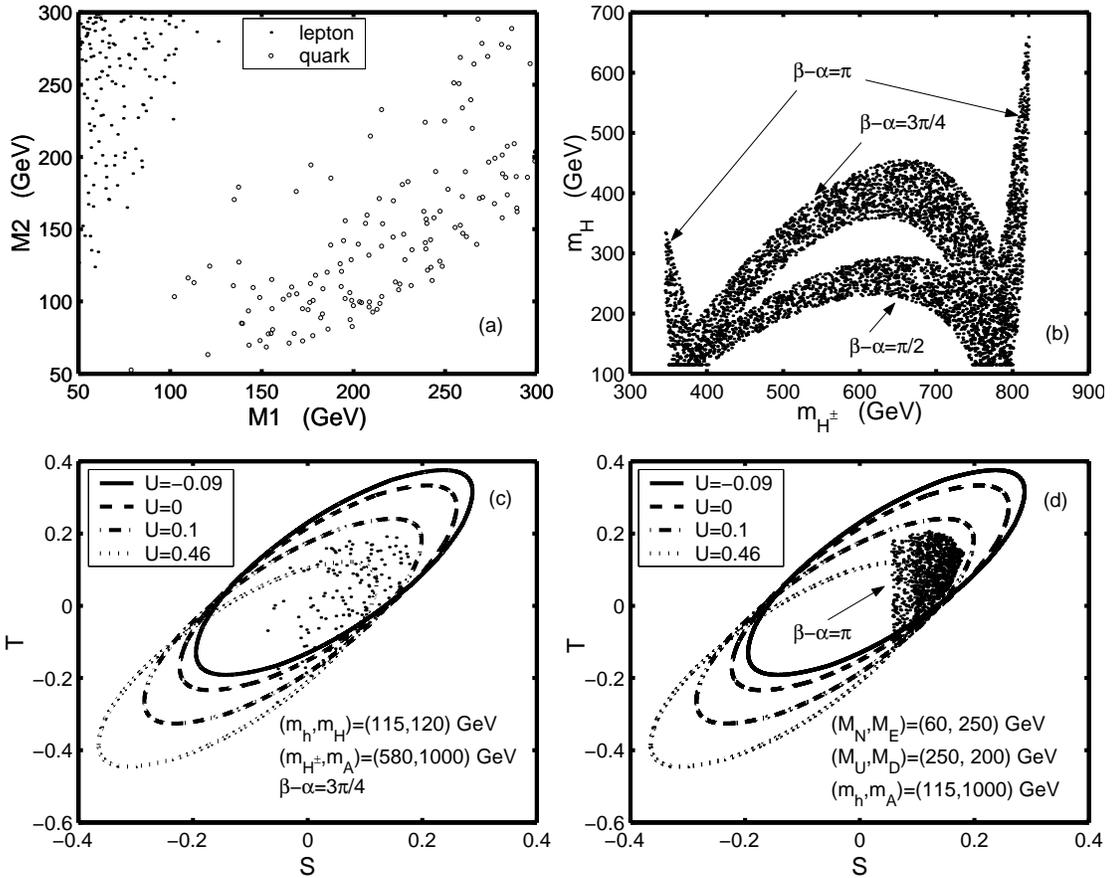}
\end{center}
\caption{
The allowed parameter space for the scenario with
three extra families and the two Higgs doublets is shown, 
(a) in the $M_{1}-M_{2}$ plane for
$(m_h, \,m_H, \,m_A, \,m_{H^\pm})= $ $(115, \,120, \,1000, \,580)$\,GeV
and $\beta-\alpha=3\pi/4$ [where dots (circles) denote leptons (quarks)];
(b)  in the Higgs mass plane 
for $(M_N, \,M_E)=(60, \,250)$ GeV, $(M_U, \,M_D)=(250, \,200)$\,GeV,
and for three values of $\beta-\alpha$. In (c) and (d),
the allowed points are shown in the $S - T$ plane
(with the $\95CL$ experimental bounds displayed),
corresponding to the inputs of (a) and (b), respectively.
For (d), only $\beta -\alpha=\pi$ is chosen for the illustration.
}
\label{fig:threeg}
\end{figure}

The two-Higgs-doublet
sector can lead to a large negative $T_H$ (cf. Sec.\,\ref{sec:2hdm}) which
will cancel to a large part the three-family fermionic $T_f$, and 
render the sum $T=T_f+T_H$ consistent with the experimental bounds
over certain regions  of the parameter space.  
For simplicity, we will assume the second and third families to
have the same mass spectrum as the first family.
This interplay is explored in
Fig.\,\ref{fig:threeg}, which is  based on an initial sample
of 50000 models. Allowed models are determined by imposing
the $\95CL$ bounds of $\STU$. 
Figs.\,\ref{fig:threeg}(a) and (b) display
the extra fermions and the Higgs bosons spectra, respectively.
We choose, for illustration,
a typical set of Higgs inputs 
$(m_h,\,m_H,\,m_A,\,m_{H^\pm}) = (115,\,120,\,1000,\,580)$\,GeV
and $\beta-\alpha=3\pi/4$ in (a), 
and a  set of fermionic inputs
$(M_N,\,M_E)=(60,\,250)$\,GeV, $(M_U,\,M_D)=(250,\,200)$\,GeV in (b),
where three values of $\beta-\alpha$ are shown.
Figs.\,\ref{fig:threeg}(c) and (d) display
the allowed points, with the same inputs as (a) and (b), respectively, 
in the $S-T$ plane for comparison
with the experimental bounds.
Variation of $m_h=115$\,GeV in the $\sim\! 100-200$\,GeV range 
does not change the results. Also, for clarity, only 
$\beta-\alpha =\pi$ is shown in (d), but similar results are obtained for
the case of $\beta-\alpha =\pi/2$, or, $3\pi/4$.

The choice of the above Higgs inputs in Figs.\,\ref{fig:threeg}(a) and (c)
corresponds to a small allowed region
in the $M_1 - M_2$ plane, i.e.,
the mirror leptons (dots) are highly non-degenerate, 
while the mirror quarks (circles) exhibit much smaller
isospin breaking.  Similar results could be obtained for other 
choices of Higgs masses, where the Higgs contribution $T_H$ is
sizable and negative.  
From Fig.\,\ref{fig:threeg}(b), one observes that the allowed regions 
are quite distinct for three choices of 
$\beta -\alpha \in (\pi/2,\,3\pi/4,\,\pi)$.
For $\beta -\alpha=\pi$, $m_H$ could vary in a wide range
[corresponding to case (Ia)] for $m_{H^\pm}\sim 800$\,GeV. 
In all other cases,  
the heavier neutral Higgs $H^0$ has to be generally much lighter than 1\,TeV. 
It is interesting to note that for
$m_{H^0} \lesssim 200$\,GeV (i.e., slightly heavier than $h^0$),
the charged Higgs mass $m_{H^\pm}$ is confined into two very
narrow regions around either $350-450$\,GeV or $750-800$\,GeV,
for a sizable range of $\beta-\alpha$.
Finally, Figs.\,\ref{fig:threeg}(c) and (d)
indicate that the relevant viable parameter space typically corresponds to 
$0\lesssim S \lesssim 0.2$ and $-0.1 \lesssim T \lesssim 0.2$.
In comparison with the scenario of
one-generation and one-Higgs-doublet
[cf., Fig.\,\ref{fig:oneg}(c)-(d)], 
the  viable region in the $S-T$ plane of Fig.\,\ref{fig:threeg}(c)-(d)
has a smaller  $T=T_f+T_H$. This
is due to the more negative $T_H$ values contributed by the
two-Higgs-doublet sector.

Clearly, there are strong correlations among the allowed
Higgs and the fermion mass ranges in the $N_g=3$ scenario.
This renders the model highly restrictive in its parameter
space, and it is thus instructive and encouraging for the relevant experimental
tests at the upcoming colliders, such as the Tevatron Run-II, the LHC and
the future lepton colliders. 
Collider signatures, however, merit 
a dedicated study and will not be discussed here.
Before concluding this subsection, we note that in the above 
we did not address explicitly the 
less difficult case of $N_g<3$.
We  expect that $N_g=1,\,2$ can be accommodated over larger
regions of the 2HDM parameter space.

\section{Conclusions}
\label{sec:sum}

In summary, we have demonstrated that one extra generation of
relatively light non-degenerate chiral fermions in the mass range, 
$m_Z/2 \lesssim M_f \lesssim \CO(\langle H \rangle)$, 
can be consistent with current precision electroweak data without
requiring additional new physics source. Sizable mass splitting
between up- and down-type fermions can lead to a 
large positive $T$ without significantly increasing  $S>0$.
This can largely relax the upper bound from precision data on the mass of 
a SM-like Higgs boson, as
shown in Fig.\,\ref{fig:oneg}.

The case of  three extra chiral families was shown to be viable
when invoking extra new physics, most notably,
a two-Higgs-doublet extension. 
In order to remain  model-independent,
we performed the analysis for three extra families with a
general two-Higgs-doublet sector. We found,  after imposing the oblique
precision bounds, a highly restrictive mass spectrum for 
either the fermion sector or
Higgs sector (cf., Fig.\,\ref{fig:threeg}),
which can lead to various distinct collider signatures.
The importance  of the two-Higgs-doublet sector is in providing
a negative contribution to $T$, and thus allowing for
a large isospin violation 
in the three family fermion sector.

We have used 
weak-scale $N=2$ supersymmetry\,\cite{N2old,N2new}
as an explicit theory framework to motivate our study and  to define
the relevant mass range for the extra chiral families under 
consideration [cf., (\ref{eq:mirrorF-MR})], as well as to
define the Higgs sector. 
We note that such an effective four-dimensional
$N=2$ structure can be a consequence of the compactification
of certain extra-dimensional theories.

Possible extensions of our study may 
include: $(i)$ a more exhaustive parameter scan 
of the two-Higgs-doublet sector, 
allowing for flavor-dependent fermion masses and family mixings;
$(ii)$ an extended Higgs sector with more than two doublets 
generating EWSB,
which is possible in $N=2$ theories\,\cite{N2new};
$(iii)$ oblique corrections from relatively light
sfermions (and mirror sfermions) and 
Majorana fermions such as gauginos, Higgsinos, and their mirrors; and 
$(iv)$ the considerations of $Z - Z^{\prime}$ mixing
in extra $U(1)^{\prime}$ models\,\cite{extraZ}. Each of these extensions
can affect, in principle, the constraints
on $N_{g}$, the two-Higgs-doublet spectrum, and their correlations.
However, these are highly 
model-dependent avenues  which are left for future works.
In addition, our study may be further extended for a six
parameter analysis including $(S,\,T,\,U,\,V,\,W,\,X)$ \cite{VWX} together,
which may be relevant for the region of $M_f \lesssim m_{Z}$.

\acknowledgements

It is our pleasure to thank Jens Erler for 
various discussions on precision data and
for his comments on the manuscript.
We also thank Howard E. Haber for conversations on the 
oblique corrections in the two-Higgs-doublet model and 
Duane A. Dicus for discussions.
H.J.H. is supported by the US Department of Energy (DOE) 
under grant DE-FG03-93ER40757;
N.P. is supported by the DOE under cooperative research 
agreement No.~DF--FC02--94ER40818; and   
S.S. is supported by the DOE under grant DE-FG03-92-ER-40701 and 
by the John A. McCone Fellowship.
\\
\\

\appendix

\section{Higgs Contributions to Oblique Parameters}

We consider general 2HDM where the Higgs bosons
$(h^0,\,H^0,\,A^0,\,H^\pm)$ have masses 
$(m_h,\,m_H,\,m_A,\,m_{H^\pm})$, respectively.
After subtracting the SM Higgs corrections to $\STU$ with reference choice
$(m_{H}^{\rm sm})_{\rm ref}=m_h$, the one-loop Higgs contributions to $\STU$
read\,\cite{haber},
\begin{eqnarray}
S_H&=&\frac{1}{\pi{m}_Z^2}\left\{\sin^2(\beta-\alpha)
{\cal B}_{22}(m_Z^2; m^2_H,m^2_A)-
{\cal B}_{22}(m_Z^2; m^2_{H^{\pm}},m^2_{H^{\pm}})\right.\nonumber \\
&&+\left.\cos^2(\beta-\alpha)\left[{\cal B}_{22}(m^2_Z; m^2_h,m^2_A)+
{\cal B}_{22}(m^2_Z;m^2_Z,m^2_H)-{\cal B}_{22}(m^2_Z;m^2_Z,m^2_h)
\right.\right.\nonumber \\
&&-\left.\left.m^2_Z{\cal B}_0(m^2_Z;m^2_Z,m^2_H)+m^2_Z
{\cal B}_0(m^2_Z;m^2_Z,m^2_h)\right]\right\},
\label{eq:Shiggs}\\
T_H&=&\frac{1}{16\pi{m}^2_Ws^2_W}
\left\{F(m^2_{H^{\pm}},m^2_A)+\sin^2(\beta-\alpha)\left[
F(m^2_{H^{\pm}},m^2_H)-F(m^2_A,m^2_H)\right]\right.\nonumber\\
&&+ \cos^2(\beta-\alpha)\left[F(m^2_{H^{\pm}},m^2_h)-F(m^2_A,m^2_h)
                             +F(m^2_W,m^2_H)-F(m^2_W,m^2_h)\right. 
\nonumber\\
&&   \left.\left.
           -F(m^2_Z,m^2_H)+F(m^2_Z,m^2_h)
           +4m^2_Z \ov{B}_0(m^2_Z,m^2_H,m_h^2)
           -4m^2_W \ov{B}_0(m^2_W,m^2_H,m_h^2)   \]\right\},
\label{eq:Thiggs}\\
U_H&=&-S_H+\frac{1}{\pi{m_Z}^2}\left\{{\cal B}_{22}(m_W^2;m^2_A,m^2_{H^{\pm}})
-2{\cal B}_{22}(m^2_W;m^2_{H^{\pm}},m^2_{H^{\pm}})\right.\nonumber \\
&&+\left.\sin^2(\beta-\alpha){\cal B}_{22}(m^2_W;m^2_H,m^2_{H^{\pm}})
+\cos^2(\beta-\alpha)\left[{\cal B}_{22}(m^2_W;m^2_h,m^2_{H^{\pm}})
\right.\right.\nonumber \\
&&+\left.\left.{\cal B}_{22}(m^2_W;m^2_W,m^2_H)
-{\cal B}_{22}(m^2_W;m^2_W,m^2_h)\right.\right.\nonumber \\
&&-\left.\left.m^2_W{\cal B}_0(m^2_W;m^2_W,m^2_H)
+m^2_W{\cal B}_0(m^2_W;m^2_W,m^2_h)\right]\right\},
\label{eq:Uhiggs}
\end{eqnarray}
where we have explicitly worked out the finite part of ${\cal B}$-functions:
\begin{eqnarray}
{\cal B}_{0}(q^2;m^2_1,m^2_2)&=&1+\frac{1}{2}\left[
\frac{x_1+x_2}{x_1-x_2}-(x_1-x_2)\right]\ln\frac{x_1}{x_2}
+\frac{1}{2}f(x_1,x_2)\,,   \\[1.5mm]
&\dis\stackrel{m_1=m_2}{\Longrightarrow}&
\dis 2 -2\sqrt{4x_1-1}\arctan\f{1}{\sqrt{4x_1-1}} \,,
\\[2mm]
\ov{B}_0(m_1^2,m_2^2,m_3^2) 
&\equiv& B_{0}(0;m^2_1,m^2_2)-B_{0}(0;m^2_1,m^2_3)
\nonumber\\[1.5mm]
&=&
\frac{m^2_1\ln{m^2_1}-m^2_3\ln{m^2_3}}{m^2_1-m^2_3}-
\frac{m^2_1\ln{m^2_1}-m^2_2\ln{m^2_2}}{m^2_1-m^2_2} \,,\\[2mm]
{\cal B}_{22}(q^2;m^2_1,m^2_2)
&\equiv& B_{22}(q^2;m^2_1,m^2_2)-B_{22}(0;m^2_1,m^2_2) \nonumber\\
&=&\frac{q^2}{24}\left\{
2\ln{q^2}+\ln(x_1x_2)
+\left[(x_1-x_2)^3-3(x_1^2-x_2^2)
\right.\right.\nonumber \\
&&+\left.\left.3(x_1-x_2)\right]\ln\frac{x_1}{x_2}
-\left[2(x_1-x_2)^2-8(x_1+x_2)+\frac{10}{3}\right]
\right.\nonumber \\
&&-\left.\left[(x_1-x_2)^2-2(x_1+x_2)+1\right]f(x_1,x_2)
-6F(x_1,x_2)\right\},\\
&\dis\stackrel{m_1=m_2}{\Longrightarrow}&
\frac{q^2}{24}\left[2\ln{q^2} + 2\ln{x_1}
+\left(16x_1-\frac{10}{3}\right)+
\left(4x_1-1\right)G(x_1)\right],  
\end{eqnarray}
\begin{eqnarray}
F(x_1,x_2)&=&\dis\frac{x_1+x_2}{2}-\frac{x_1x_2}{x_1-x_2}\ln\frac{x_1}{x_2}
\,,
\label{eq:Ffun}
\\[2mm]
G(x)&=&\dis -4\sqrt{4x-1}\,\arctan\frac{1}{\sqrt{4x-1}}
\,,
\label{eq:Gfun}
\\[2mm]
f(x_1,x_2)&=&\left\{
\begin{array}{ll}
-2\sqrt{\Delta}\left[\dis\arctan\frac{x_1-x_2+1}{\sqrt{\Delta}}
-\arctan\frac{x_1-x_2-1}{\sqrt{\Delta}}\right]\,,
&~~(\Delta>0)\,,  \\[1.5mm] 
0\,,
&~~(\Delta=0)\,,\\[1.5mm]
\sqrt{-\Delta}\dis\ln\frac{x_1+x_2-1+\sqrt{-\Delta}}{x_1+x_2-1-\sqrt{-\Delta}}\,,
&~~(\Delta<0)\,,
\end{array}
\right. 
\label{eq:ffun}\\[3mm]
\Delta&=&2(x_1+x_2)-(x_1-x_2)^2-1  \,,
\end{eqnarray}
with ~$\dis x_i \equiv \f{m_i^2}{q^2}$\,.

The various expressions are simplified in
the limit of $m_{\rm Higgs}^2\gg{m}_Z^2$. The approximate formula for 
$T_H$ in this limit has already been given in eq.~(\ref{eq:thiggsapp}).  
Similarly, eqs.~(\ref{eq:Shiggs}) and (\ref{eq:Uhiggs}) 
reduce in this limit to 
\begin{eqnarray}
S_H&=&\frac{1}{12\pi}\left(\cos^2(\beta-\alpha)\[
\ln\frac{m_H^2}{m_h^2}+g(m_h^2,m_A^2)-\ln\frac{m_{H^{\pm}}^2}{m_hm_A}\]
\right.\nonumber \\
&+&\left.\sin^2(\beta-\alpha)\[g(m_H^2,m_A^2)-\ln\frac{m_{H^{\pm}}^2}{m_Hm_A}
\]\right),\\
U_H&=&\frac{1}{12\pi}\left(\cos^2(\beta-\alpha)\[g(m_h^2,m_{H^{\pm}}^2)
+g(m_A^2,m_{H^{\pm}}^2)-g(m_h^2,m_A^2)\]
\right.\nonumber \\
&+&\left.\sin^2(\beta-\alpha)\[g(m_H^2,m_{H^{\pm}}^2)+g(m_A^2,m_{H^{\pm}}^2)
-g(m_H^2,m_A^2)\]\right),
\end{eqnarray}
where
\begin{equation}
g(x_1,x_2)=-\frac{5}{6}+\frac{2x_1x_2}{(x_1-x_2)^2}+
\frac{(x_1+x_2)(x_1^2-4x_1x_2+x_2^2)}{2(x_1-x_2)^3}\ln\frac{x_1}{x_2}.
\end{equation}


\end{document}